\documentclass[prd,onecolumn,preprint,nofootinbib]{revtex4}
\pdfoutput=1
\usepackage{amssymb}
\usepackage{graphicx}
\newcommand{\rthis}[1]{\textcolor{black}{#1}}

\usepackage{lipsum}

\usepackage[plainpages=false, colorlinks=true, anchorcolor=blue, linkcolor=blue, citecolor=blue, allcolors=blue, bookmarks=false]{hyperref}
\usepackage{textgreek}
\usepackage[utf8]{inputenc}
\usepackage[english]{babel}
\usepackage{amsmath}
\usepackage{hyperref}
\usepackage{float}

\begin{document}
\title{Search for Lorentz Invariance Violation with  spectral lags of GRB 190114C using profile likelihood}

\author{Vyaas Ramakrishnan}
\email{vyaas3305@gmail.com}

\author{Shantanu Desai}
\email{shntn05@gmail.com}

\begin{abstract}
  We search for Lorentz invariance violation (LIV) by re-analyzing the spectral lag data for GRB 190114C \rthis{from Fermi-GBM} using frequentist analysis, where we deal with the astrophysical nuisance parameters using profile likelihood. For this use case, we find a global minima for the $\chi^2$ as a function of   energy scale of LIV ($E_{QG}$), well  below the Planck scale. The best-fit $1\sigma$ central intervals for $E_{QG}$  are given by $2.81^{+0.50}_{-0.37}\times 10^{14}$ GeV and $9.85^{+0.84}_{-0.60}\times 10^{5}$ GeV for linear and quadratic LIV, respectively, and agree with the Bayesian estimates obtained so far in a previous work. Therefore, the results from frequentist analysis GRB 190114C  agrees with Bayesian analysis. 
\end{abstract}

\affiliation{Department of Physics, Indian Institute of Technology, Hyderabad, Kandi Telangana 502284, India}
\maketitle

\section{Introduction}
\label{sec:intro}
The spectral lags of gamma-ray bursts (GRB) have proved to be a powerful probe of Lorentz invariance violation (LIV)~\citep{Desairev,WuGRBreview,WeiWu2}. The spectral lag is defined as the time difference between the arrival of high- and low-energy photons. Until recently, all searches for LIV using GRB spectral lags \rthis{over a multi-band energy range  have  used Bayesian analysis to marginalize over the astrophysically induced spectral lags}.    A  frequentist  search for LIV using  the spectral lag data for GRB 160625B  was first carried out  in ~\cite{Ganguly24}, where the astrophysical nuisance parameters were dealt with using profile likelihood, instead of marginalization as is done in Bayesian analysis. This was followed by  another frequentist-based search for LIV with the spectral lags of 56 GRBs in the cosmological rest frame~\citep{Vyaas25}. \rthis{We note however that multiple searches for LIV using GRB spectral lags with single photons  have been done using frequentist analysis~\cite{Martinez08,Vaso13,MAGIC}.}

In this work, we carry out yet another test of \rthis{sub-luminal} LIV with frequentist analysis using the  spectral lag data for GRB 190114C, which  hitherto has  been analyzed using Bayesian inference~\citep{Du20}. This manuscript is structured as follows. The analysis methodology is described in Sect.~\ref{sec:analysis}. Our results are discussed in Sect.~\ref{sec:results} and we  conclude  in Sect.~\ref{sec:conclusions}.

\section{Data and Analysis}
\label{sec:analysis}
GRB 190114C is a long GRB, which was triggered by the \rthis{on-board software}  on SWIFT-BAT and Fermi-GBM on 14 Jan. 2019, \rthis{due to a statistically significant increase in gamma-ray counts.}
This GRB is located at a redshift $z=(0.4245 \pm 0.0005)$. \citet{Du20} (D20, hereafter) have calculated the GRB spectral lags with respect to the reference energy of 10-12 keV based on the cross-correlation method~\citep{Ukwatta,Zhang12} using data from Fermi-GBM. \rthis{The cross-correlation method is the most robust among existing techniques when the GRB light curve is sufficiently complex  and not dominated by a prominent pulse~\cite{Ukwatta}.}
D20 obtained 19 spectral lags with respect to the lowest energy band of (10-15) keV, with the highest energy band equal to (1700-5000) keV. These spectral lags have been collated in Table 1 of D20.  We note the   spectral lag data for this GRB  was also been combined with the same from other GRBs in a stacked search for LIV in multiple works~\cite{Agrawal_2021,Liu22,Desai23,Pasumarti23}. These works had found a \rthis{best-fit value for the energy scale for LIV below the Planck scale}. Results for a search for anisotropic LIV using the spectral lag data for this GRB have also been reported in ~\cite{Weiuniverse}.

 The expected  spectral time lag from this GRB can be written as the sum of the intrinsic astrophysical lag and that induced because of LIV.
\begin{equation}
\Delta t_{exp} = (1+z) \Delta t_{int} + \Delta t_{LIV},
\label{eq:sum}
\end{equation}
where $\Delta t_{int}$ is the intrinsic time lag in the GRB rest frame due to astrophysical effects and $z$ is the redshift.  D20 used the following parameterization for the intrinsic time lag, which has also been previously used to model the time lags of GRB 160625B~\citep{Wei}.
\begin{equation}
\Delta t_{int} (E)   = \tau \left[\left(\frac{E_0}{keV}\right)^{-\alpha} -  \left(\frac{E}{keV}\right)^{-\alpha} \right], 
\label{eq:int}
\end{equation}
where $E_0=12.5$ keV. \rthis{The parameters $\tau$ and $\alpha$ are free parameters for this power-law model, corresponding to the   scaling factor and exponent, respectively. This particular power-law model was motivated from a study of light curves of about 50 GRBs which showed that the spectral time lags are correlated with energy~\cite{Shao16}.}
In Eq.~\ref{eq:sum}, the expression for sub-luminal LIV-induced time lag ($\Delta t_{LIV}$) is given by~\citep{Jacob}:
\begin{equation}
\Delta t_{LIV} =  -\left(\frac{1+n}{2H_0}\right)\left(\frac{E^n - E_0^n}{E_{QG,n}^n}\right)\frac{1}{(1+z)^n}\int_{0}^{z} \frac{(1+z^{\prime})^n}{h(z^{\prime})} \, dz^{\prime} 
\label{eq:LIV}
\end{equation}
where $E_{QG,n}$ is the quantum gravity scale corresponding to the scale of LIV and $H_0$ is the Hubble constant. In Eq.~\ref{eq:LIV}, $n=1$ and  $n=2$ correspond to  linear  and  quadratic LIV models, respectively. 
In Eq.~\ref{eq:LIV},
$h(z) \equiv \frac{H(z)}{H_0}$ is the dimensionless  Hubble parameter as a function of redshift. For the current standard flat $\Lambda$CDM model, $h(z)= \sqrt{\Omega_M (1+z^\prime)^3 + 1-\Omega_M}$, where $\Omega_M$ is the cosmological matter density. 
  On the other hand, D20 as well as one of our previous works~\citep{Agrawal_2021} parameterized $h(z)$ in a model-independent manner with Gaussian process regression using cosmic chronometers as probes of expansion history. However, in another work on LIV using spectral lags, we have shown that the results of the two are comparable~\citep{Desai23}, and therefore we use the $\Lambda$CDM model to parameterize $h(z)$.

We now construct the maximum likelihood which is given by
\begin{equation}
     \mathcal{L}=\prod_{i=1}^N \frac{1}{\sigma_{tot,i} \sqrt{2\pi}} \exp \left\{-\frac{[\Delta t_i-\Delta t_{exp}]^2}{2\sigma_{tot,i}^2}\right\},
     \label{eq:likelihood}
  \end{equation}
where $\Delta t_i$ corresponds to the observed time lag; $\Delta t_{exp}$ is given by Eq.~\ref{eq:sum};  $N$ is the total number of spectral lags for GRB 190114C (19); and  $\sigma_{tot,i}$ denotes the total uncertainty, which is given by: 
\begin{equation} 
\sigma_{tot,i}^2 = \sigma_i^2 + \Big(\frac{\partial \Delta t_{exp}}{\partial E}\Big)^2\sigma_E^2 ,
\label{eq:totalerror}.
\end{equation}
where $\sigma_i$ is the uncertainty in the spectral delay \rthis{at each lag} and $\sigma_E$ is the uncertainty in the energy band, and as described in D20, corresponds to half of each energy band,  given by $0.5 (E_{max}-E_{min})$. \rthis{We assume that the uncertainties in each spectral lag (which have been obtained by Monte-Carlo simulations~\cite{Ukwatta}) are uncorrelated. We also note that we have assumed a Gaussian likelihood in Eq.~\ref{eq:likelihood}.  However, this assumption may not hold if the  uncertainties in spectral-lag  are not Gaussian or independent. The spectral-lag measurements derived from cross-correlations could be  non-Gaussian and correlated, particularly at high energies or for small lags. We also note that we are not using the light curve data directly, but only the energy-resolved light curves. These  lag measurements could sometimes be constrained  due to  pulse overlap~\cite{Zhang12}. More details on potential systematics in the estimation of spectral lags can be found in ~\cite{Ukwatta}.
However, our main goal is to compare with the Bayesian results in D20, which also used a  Gaussian likelihood. Therefore, we have also directly used the same uncertainties as in D20.}

The most general procedure for constructing frequentist confidence intervals involves a Neyman construction~\cite{Herold24}. \rthis{However, this procedure entails generating synthetic catalogs for three free parameters and is computationally expensive. There are also multiple methods for dealing with multiple nuisance parameters for ensuring proper coverage~\cite{Nova}. Therefore, the full Neyman construction is generally avoided and the confidence intervals are obtained using the method of graphical profile likelihood, which is usually refered to as profile likelihood. A comparison of the graphical profile likelihood with other frequentist methods has recently been discussed in ~\cite{Colgain25}.}
We  now define $\chi^2 \equiv -2 \ln \mathcal{L}$~\cite{Wilks} where $\mathcal{L}$ is defined in Eq.~\ref{eq:likelihood}. For this problem, the parameter of interest is $E_{QG}$, while $\tau$ and $\alpha$ constitute the nuisance parameters. We first construct a grid of $E_{QG}$ with a log-uniform spacing between $10^6$ GeV and $10^{19}$ GeV. For each fixed value of $E_{QG}$, we  calculate the minimum $\chi^2$ by keeping  $\tau$ and $\alpha$ as free parameters.  We then repeat this procedure for all values of $E_{QG}$  to determine the global  minimum $\chi^2$ ($\chi^2_{min}$).
According to Wilks' theorem $\Delta \chi^2$ defined as $ \chi^2 (E_{QG}) -\chi^2_{min}$ follows a $\chi^2$ distribution for one degree of freedom~\citep{Wilks}. We now describe the results of our analysis in the next section.
 \section{Results}

\label{sec:results}
\begin{figure}[H]
    \centering
    \includegraphics[width=0.7\linewidth]{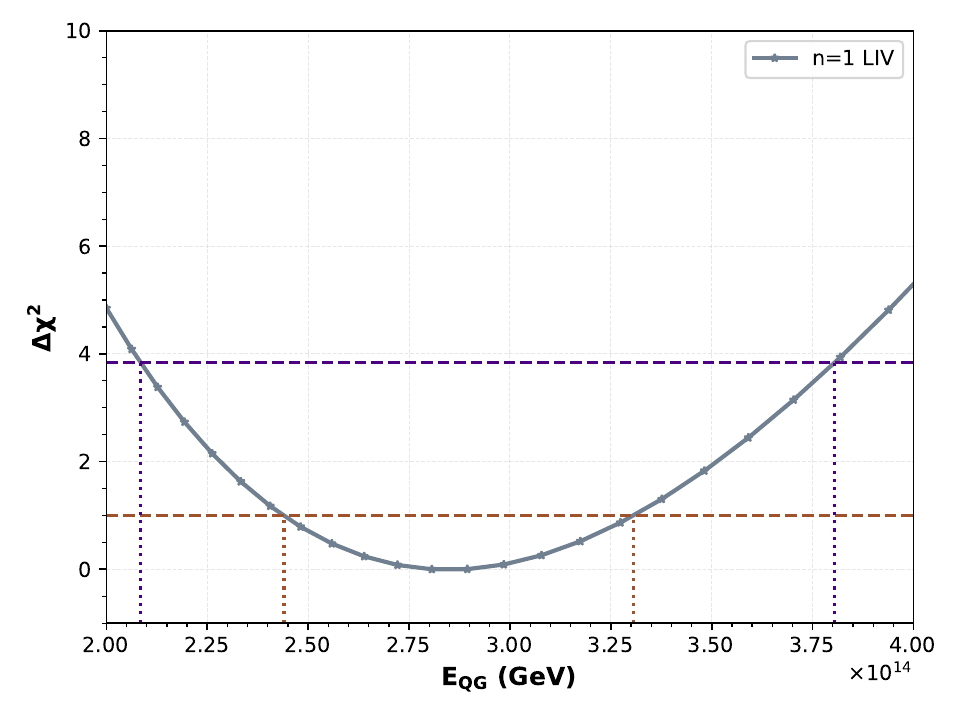}
    \caption{$\Delta \chi^2$, defined as ($\chi^2-\chi^2_{min}$), plotted against $E_{QG}$ for a linearly-dependent LIV, corresponding to $n = 1$. The horizontal magenta dashed line represents $\Delta\chi^2 = \rthis{3.84}$ and the horizontal red dashed line represents $\Delta\chi^2 = 1$. The corresponding X-intercepts, provide us the both the 68.3\% confidence interval ($\Delta\chi^2 = 1$) for $E_{QG,1} = 2.81^{+0.50}_{-0.37}\times 10^{14}$ GeV and the 95\% confidence interval \rthis{($\Delta\chi^2 = 3.84$)  for $E_{QG,1} = 2.81^{+0.99}_{-0.73}\times 10^{14}$ GeV.}}
    \label{fig:linearLIV}
\end{figure}

\begin{figure}[H]
    \centering
    \includegraphics[width=0.7\linewidth]{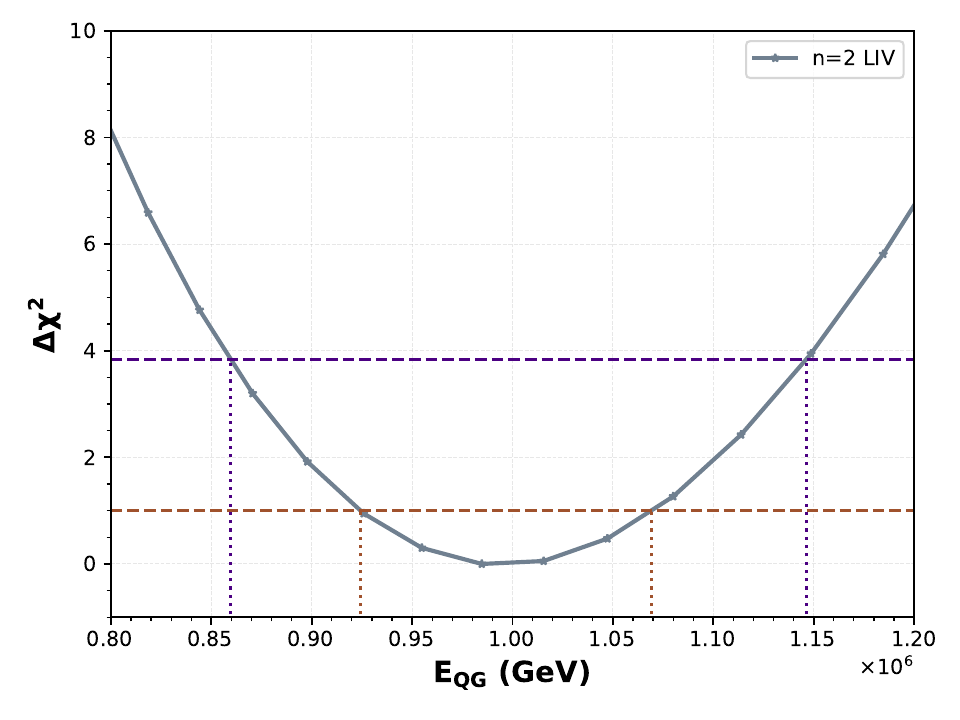}
    \caption{$\Delta \chi^2$, defined as ($\chi^2-\chi^2_{min}$), plotted against $E_{QG}$ for a quadratically-dependent LIV, corresponding to $n = 2$. The horizontal indigo dashed line represents $\Delta\chi^2 = \rthis{3.84}$ and the horizontal sienna dashed line represents $\Delta\chi^2 = 1$. The corresponding x-intercepts, provide us the both the 68.3\% confidence interval ($\Delta\chi^2 = 1$) for $E_{QG,2} = 9.85^{+0.84}_{-0.60}\times 10^{5}$ GeV and the 95\% \rthis{confidence interval ($\Delta\chi^2 = 3.84$) for $E_{QG,2} = 9.85^{+1.62}_{-1.25}\times 10^{5}$ GeV.}}
    \label{fig:quadLIV}
\end{figure}

 The $\chi^2$ minimization was done using  {\tt scipy.optimize.fmin} module in Python. We also compared these results with other minimization  algorithms implemented in {\tt scipy}, and all of them gave the same result.
For both models of LIV, we find a  parabolic shape for $\chi^2$ as a function of $E_{QG}$ with a global minimum below the Planck scale. The plots of $\Delta \chi^2$ as a function of $E_{QG}$ can be found in Fig.~\ref{fig:linearLIV} and Fig.~\ref{fig:quadLIV} for linear and quadratic LIV, respectively. The 95\% confidence level  central intervals are obtained by finding the X-intercept corresponding to $\Delta \chi^2=3.84$, and the 68.3\% confidence level are obtained from the same for $\Delta \chi^2=1.0$.
The best-fit 95\% confidence interval estimates thus obtained for $E_{QG}$ are equal to  $2.81^{+0.99}_{-0.73}\times 10^{14}$ GeV and 
$9.85^{+1.62}_{-1.25}\times 10^{5}$ for linear and quadratic LIV, respectively. The corresponding $1\sigma$ central estimates obtained by finding the X-intercept corresponding to $\Delta \chi^2=1.0$  are given by $2.81^{+0.50}_{-0.37}\times 10^{14}$ GeV and  $9.85^{+0.84}_{-0.64}\times 10^{5}$ GeV
for linear and quadratic LIV, respectively.
These values agree within $1\sigma$ compared to the marginalized 68.3\% credible intervals for $E_{QG}$ estimated in D20. \rthis{The difference between $\chi^2_{min}$ and $\chi^2_{\infty}$ (which we set to Planck scale of $10^{19}$ GeV) is equal to 368 and 365, for linear and quadratic models of LIV, respectively. These correspond to significances of about $19\sigma$ for both the models. }

In order to judge the efficacy of the fit,  we calculate the $\chi^2_{fit}$ based on the residuals between the data and best-fit model:
\begin{equation}
\chi^2_{fit} =\sum_i^N \left(\frac{\Delta t_{i}-\Delta t_{exp}}{\sigma_{tot,i}}\right)^2\, 
\label{eq:chifit}
\end{equation}
where all terms have the same meaning as in Eq.~\ref{eq:likelihood}.
Note that $\chi^2_{fit}$ is \rthis{a residual-based goodness-of-fit diagnostic} used to ascertain the quality of the fit. 
%It  differs from $\chi^2$ obtained using profile likelihood, since the latter contains an extra term $\frac{1}{\sigma_{tot}}$ outside the exponent. 
The best-fit $\chi^2$ along with the DOF can be found in Table~\ref{tab:model_comp}. Similar to D20, our best-fit reduced  $\chi^2_{fit}$ values are  $< 1$.  We find that the reduced $\chi^2$ values are less than 1, confirming that the fit is good \rthis{assuming the uncertainties are correct and not overestimated.}
Table~\ref{tab:model_comp} also contains the best-fit values for $\tau$ and $\alpha$ for both the LIV models.  These best-fit values also agree within $1\sigma$ with the corresponding marginalized 68\% credible intervals for $\tau$ and $E_{QG}$ obtained in D20.

Therefore, for GRB 190114C, the results from frequentist inference agree with those from Bayesian inference. We note however that the result from this one GRB should not be construed as evidence for LIV. Currently, the most stringent bounds on LIV is from the LHAASO observation of TeV afterglow of GRB 221009A, which yields $E_{QG} \geq 10 E_{pl}$ and $E_{QG} \geq 6 \times 10^{-8} E_{pl}$ for linear and quadratic LIV, respectively~\cite{LHAASO}. There are also multiple results using the spectral data from GRB 090510 and other GRBs~\cite{Fermi09,Abdo09,Vaso13},  which rule out LIV at  energy scales scales greater than the $1\sigma$ central interval obtained in this work and D20. A comprehensive summary of all other limits
can be found elsewhere~\cite{Desairev,WuGRBreview,WeiWu2}.
We also note that GRB 190114C was detected above 0.2 TeV by the MAGIC  telescope~\cite{MAGICGRB}. No evidence of LIV was found using the spectral lags observed by MAGIC  and lower limits on $E_{QG}$ were established,  which are given by $E_{QG} \gtrapprox 0.6 \times 10^{19}$ GeV and $E_{QG} \gtrapprox 6 \times 10^{10}$ GeV for linear and quadratic LIV, respectively~\cite{MAGIC}.  The MAGIC data for this GRB has also been combined with TeV spectral lags from other GRBs to search  for LIV~\cite{Song25,Song2}. 

Therefore, the lower limits on $E_{QG}$ using  TeV gamma-ray data from GRB 190114C as well as from multiple  other GRBs  rule out the $1\sigma$ central estimate   we have found in this work and D20. So the spectral lag data for this GRB in the keV energy range for  GRB 190114C is most likely due to some intrinsic astrophysical mechanism and cannot be due to LIV. We note however that our main goal here was only to compare the frequentist parameter estimates with the Bayesian estimates in D20,  and a detailed discussion of the possible astrophysical mechanisms which cause the spectral lag is beyond the scope of this work.

\begin{table}[H]
    \centering
    \renewcommand{\arraystretch}{1.5}
    \begin{tabular}{|c|cc|cc|}
         \hline
         {} & \multicolumn{2}{c|}{\textbf{Linear LIV}} & \multicolumn{2}{c|}{\textbf{Quadratic LIV}}\\
         {} & \multicolumn{2}{c|}{$\mathbf{n=1}$} & \multicolumn{2}{c|}{$\mathbf{n=2}$}\\
         \hline
         $\tau$ & \multicolumn{2}{c|}{$5.24$} & \multicolumn{2}{c|}{$10.56$}\\
         $\alpha$ & \multicolumn{2}{c|}{$0.67$} & \multicolumn{2}{c|}{$1.04$}\\
         $E_{QG}$ (GeV) & \multicolumn{2}{c|}{$2.81 \times 10^{14}$} & \multicolumn{2}{c|}{$9.85 \times 10^{5}$}\\
         \hline
         {$\chi^2_{fit}$}/DOF & \multicolumn{2}{c|}{$5.46/16$} & \multicolumn{2}{c|}{$5.09/16$}\\
         \hline
    \end{tabular}
    \caption{Best-fit model parameters: $\tau$ and $\alpha$, corresponding to both linear and quadratic LIV models, evaluated at their best-fit values of $E_{QG}$. The standard frequentist goodness-of-fit metric, $\chi^2_{fit}$/DOF (refer Eq.\ref{eq:chifit}), is also reported at the best-fit $E_{QG}$ values. Here, DOF refers to the degrees of freedom, which is equal to the difference between total number of data points and number of free parameters.}
    \label{tab:model_comp}
\end{table}

\section{Conclusions}
\label{sec:conclusions}
In this work, we \rthis{used} the spectral lag data of GRB 190114C \rthis{collated in D20 over the energy range from 15-5000 keV},  in order to search for \rthis{subluminal} LIV using frequentist inference, which is complementary to the Bayesian analysis carried out in D20. \rthis{We note that this multi-band search for LIV is complementary to some of the earlier searches done with single photons. }
For this purpose, \rthis{we used profile likelihood, which profiles over the   astrophysical nuisance  parameters (related to the intrinsic emission) by maximizing the likelihood with respect to their values for each fixed $E_{QG}$. }

In this case, we find a global minimum for $\chi^2$ as a function of $E_{QG}$ below the Planck scale for both the LIV models.  \rthis{These correspond to about $19\sigma$ improvement over the no-LIV (for $E_{QG}=10^{19}$ GeV) scenario.}
The $\Delta \chi^2$ \rthis{curves (profile likelihood scans)}  as a function of $E_{QG}$ can be found in Fig.~\ref{fig:linearLIV} and  Fig.~\ref{fig:quadLIV} for linear and quadratic LIV, respectively. The \rthis{two-sided confidence} intervals (68.3\%) for  $E_{QG}$ are given by   $2.81^{+0.50}_{-0.37}\times 10^{14}$ GeV and 
$9.85^{+0.84}_{-0.60}\times 10^{5}$ for linear and quadratic LIV, respectively. \rthis{We note however these confidence intervals assume the validity of Wilks' theorem, which has not been verified for this dataset.}
The best-fit  values agree within $1\sigma$ compared to the marginalized 68\% credible intervals for $E_{QG}$ estimated in D20. Therefore, for the spectral lag data of GRB 190114C,  the results of frequentist analysis agree with Bayesian inference. \rthis{However, this result should not be construed as evidence for LIV, since the best-fit values for $E_{QG}$ are ruled out by lower limits from other searches for LIV~\cite{Fermi09,Abdo09,Vaso13,MAGIC}.}  This work mainly represents  a  proof of principles application of profile likelihood in searches for Lorentz invariance violation
using GRB spectral lags.

In the spirit of open science, we have made our analysis codes publicly available on Github, which can be found \href{https://github.com/vyaas3305/liv-PL-GRB190114C}{here}.

\section*{Acknowledgements}
\rthis{We are thankful to the anonymous referee for many useful comments and feedback on our manuscript.}

\bibliography{main}

@ARTICLE{Ukwatta,
       author = {{Ukwatta}, T.~N. and {Stamatikos}, M. and {Dhuga}, K.~S. and {Sakamoto}, T. and {Barthelmy}, S.~D. and {Eskandarian}, A. and {Gehrels}, N. and {Maximon}, L.~C. and {Norris}, J.~P. and {Parke}, W.~C.},
        title = "{Spectral Lags and the Lag-Luminosity Relation: An Investigation with Swift BAT Gamma-ray Bursts}",
      journal = {\apj},
         year = 2010,
        month = mar,
       volume = {711},
       number = {2},
        pages = {1073-1086},
          doi = {10.1088/0004-637X/711/2/1073},
archivePrefix = {arXiv},
       eprint = {0908.2370},
 primaryClass = {astro-ph.HE},
       adsurl = {https://ui.adsabs.harvard.edu/abs/2010ApJ...711.1073U}
}

@ARTICLE{Zhang12,
       author = {{Zhang}, Bin-Bin and {Burrows}, David N. and {Zhang}, Bing and {M{\'e}sz{\'a}ros}, Peter and {Wang}, Xiang-Yu and {Stratta}, Giulia and {D'Elia}, Valerio and {Frederiks}, Dmitry and {Golenetskii}, Sergey and {Cummings}, Jay R. and {Norris}, Jay P. and {Falcone}, Abraham D. and {Barthelmy}, Scott D. and {Gehrels}, Neil},
        title = "{Unusual Central Engine Activity in the Double Burst GRB 110709B}",
      journal = {\apj},
         year = 2012,
        month = apr,
       volume = {748},
       number = {2},
          eid = {132},
        pages = {132},
          doi = {10.1088/0004-637X/748/2/132},
archivePrefix = {arXiv},
       eprint = {1111.2922},
 primaryClass = {astro-ph.HE},
       adsurl = {https://ui.adsabs.harvard.edu/abs/2012ApJ...748..132Z}
}

@article{Wilks,
  title={The large-sample distribution of the likelihood ratio for testing composite hypotheses},
  author={Wilks, Samuel S},
  journal={The annals of mathematical statistics},
  volume={9},
  number={1},
  pages={60--62},
  year={1938},
  publisher={JSTOR}
}

@INCOLLECTION{WuGRBreview,
       author = {{Yu}, Yun-Wei and {Gao}, He and {Wang}, Fa-Yin and {Zhang}, Bin-Bin},
        title = "{Gamma-Ray Bursts}",
    booktitle = {Handbook of X-ray and Gamma-ray Astrophysics. Edited by Cosimo Bambi and Andrea Santangelo},
         year = 2022,
          eid = {31},
        pages = {31},
          doi = {10.1007/978-981-16-4544-0_126-1},
       adsurl = {https://ui.adsabs.harvard.edu/abs/2022hxga.book...31Y}
}

@INCOLLECTION{Desairev,
       author = {{Desai}, Shantanu},
        title = "{Astrophysical and Cosmological Searches for Lorentz Invariance Violation}",
    booktitle = {Recent Progress on Gravity Tests. Challenges and Future Perspectives},
         year = 2024,
       editor = {{Bambi}, Cosimo and {C{\'a}rdenas-Avenda{\~n}o}, Alejandro},
        pages = {433-463},
          doi = {10.1007/978-981-97-2871-8_11},
       adsurl = {https://ui.adsabs.harvard.edu/abs/2024rpgt.book..433D}
}

@ARTICLE{Wei,
   author = {{Wei}, J.-J. and {Zhang}, B.-B. and {Shao}, L. and {Wu}, X.-F. and 
   {M{\'e}sz{\'a}ros}, P.},
    title = "{A New Test of Lorentz Invariance Violation: The Spectral Lag Transition of GRB 160625B}",
  journal = {\apjl},
archivePrefix = "arXiv",
   eprint = {1612.09425},
 primaryClass = "astro-ph.HE",
     year = 2017,
    month = jan,
   volume = 834,
      eid = {L13},
    pages = {L13},
      doi = {10.3847/2041-8213/834/2/L13},
   adsurl = {http://adsabs.harvard.edu/abs/2017ApJ...834L..13W}
}

@ARTICLE{Shao16,
       author = {{Shao}, Lang and {Zhang}, Bin-Bin and {Wang}, Fu-Ri and {Wu}, Xue-Feng and {Cheng}, Ye-Hao and {Zhang}, Xi and {Yu}, Bang-Yao and {Xi}, Bao-Jia and {Wang}, Xue and {Feng}, Huan-Xue and {Zhang}, Meng and {Xu}, Dong},
        title = "{A New Measurement of the Spectral Lag of Gamma-Ray Bursts and its Implications for Spectral Evolution Behaviors}",
      journal = {\apj},
         year = 2017,
        month = aug,
       volume = {844},
       number = {2},
          eid = {126},
        pages = {126},
          doi = {10.3847/1538-4357/aa7d01},
archivePrefix = {arXiv},
       eprint = {1610.07191},
 primaryClass = {astro-ph.HE},
       adsurl = {https://ui.adsabs.harvard.edu/abs/2017ApJ...844..126S}
}

@INCOLLECTION{WeiWu2,
       author = {{Wei}, Jun-Jie and {Wu}, Xue-Feng},
        title = "{Tests of Lorentz Invariance}",
    booktitle = {Handbook of X-ray and Gamma-ray Astrophysics. Edited by Cosimo Bambi and Andrea Santangelo},
         year = 2022,
          eid = {82},
        pages = {82},
          doi = {10.1007/978-981-16-4544-0_132-1},
       adsurl = {https://ui.adsabs.harvard.edu/abs/2022hxga.book...82W}
}

@ARTICLE{Pasumarti23,
       author = {{Pasumarti}, Vibhavasu and {Desai}, Shantanu},
        title = "{Bayesian evidence for spectral lag transition due to Lorentz invariance violation for 32 Fermi/GBM Gamma-ray bursts}",
      journal = {Journal of High Energy Astrophysics},
         year = 2023,
        month = nov,
       volume = {40},
        pages = {41-48},
          doi = {10.1016/j.jheap.2023.10.001},
archivePrefix = {arXiv},
       eprint = {2307.02296},
 primaryClass = {astro-ph.HE},
       adsurl = {https://ui.adsabs.harvard.edu/abs/2023JHEAp..40...41P}
}

@ARTICLE{Agrawal_2021,
       author = {{Agrawal}, Rajdeep and {Singirikonda}, Haveesh and {Desai}, Shantanu},
        title = "{Search for Lorentz Invariance Violation from stacked Gamma-Ray Burst spectral lag data}",
      journal = {\jcap},
         year = 2021,
        month = may,
       volume = {2021},
       number = {5},
          eid = {029},
        pages = {029},
          doi = {10.1088/1475-7516/2021/05/029},
archivePrefix = {arXiv},
       eprint = {2102.11248},
 primaryClass = {astro-ph.HE},
       adsurl = {https://ui.adsabs.harvard.edu/abs/2021JCAP...05..029A}
}

@ARTICLE{Desai23,
       author = {{Desai}, Shantanu and {Agrawal}, Rajdeep and {Singirikonda}, Haveesh},
        title = "{Search for Lorentz invariance violation using Bayesian model comparison applied to Xiao et al. GRB spectral lag catalog}",
      journal = {European Physical Journal C},
         year = 2023,
        month = jan,
       volume = {83},
       number = {1},
          eid = {63},
        pages = {63},
          doi = {10.1140/epjc/s10052-023-11229-z},
archivePrefix = {arXiv},
       eprint = {2205.12780},
 primaryClass = {astro-ph.HE},
       adsurl = {https://ui.adsabs.harvard.edu/abs/2023EPJC...83...63D}
}

@ARTICLE{Jacob,
   author = {{Jacob}, U. and {Piran}, T.},
    title = "{Lorentz-violation-induced arrival delays of cosmological particles}",
  journal = {\jcap},
archivePrefix = "arXiv",
   eprint = {0712.2170},
     year = 2008,
    month = jan,
   volume = 1,
      eid = {031},
    pages = {031},
      doi = {10.1088/1475-7516/2008/01/031},
   adsurl = {http://adsabs.harvard.edu/abs/2008JCAP...01..031J}
}

@ARTICLE{Ganguly24,
       author = {{Desai}, Shantanu and {Ganguly}, Shalini},
        title = "{Constraint on Lorentz invariance violation for spectral lag transition in GRB 160625B using profile likelihood}",
      journal = {European Physical Journal C},
         year = 2025,
        month = mar,
       volume = {85},
       number = {3},
          eid = {290},
        pages = {290},
          doi = {10.1140/epjc/s10052-025-14016-0},
archivePrefix = {arXiv},
       eprint = {2411.09248},
 primaryClass = {astro-ph.HE},
       adsurl = {https://ui.adsabs.harvard.edu/abs/2025EPJC...85..290D}
}

@ARTICLE{Herold24,
       author = {{Herold}, Laura and {Ferreira}, Elisa G.~M. and {Heinrich}, Lukas},
        title = "{Profile likelihoods in cosmology: When, why, and how illustrated with {\ensuremath{\Lambda}}CDM, massive neutrinos, and dark energy}",
      journal = {\prd},
         year = 2025,
        month = apr,
       volume = {111},
       number = {8},
          eid = {083504},
        pages = {083504},
          doi = {10.1103/PhysRevD.111.083504},
archivePrefix = {arXiv},
       eprint = {2408.07700},
 primaryClass = {astro-ph.CO},
       adsurl = {https://ui.adsabs.harvard.edu/abs/2025PhRvD.111h3504H}
}

@ARTICLE{Vyaas25,
       author = {{Ramakrishnan}, Vyaas and {Desai}, Shantanu},
        title = "{Constraints on Lorentz Invariance Violation from Gamma-Ray Burst Rest-Frame Spectral Lags Using Profile Likelihood}",
      journal = {Universe},
         year = 2025,
        month = jun,
       volume = {11},
       number = {6},
          eid = {183},
        pages = {183},
          doi = {10.3390/universe11060183},
archivePrefix = {arXiv},
       eprint = {2502.00805},
 primaryClass = {astro-ph.HE},
       adsurl = {https://ui.adsabs.harvard.edu/abs/2025Univ...11..183R}
}

@ARTICLE{Martinez08,
       author = {{Mart{\'\i}nez}, Manel and {Errando}, Manel},
        title = "{A new approach to study energy-dependent arrival delays on photons from astrophysical sources}",
      journal = {Astroparticle Physics},
         year = 2009,
        month = apr,
       volume = {31},
       number = {3},
        pages = {226-232},
          doi = {10.1016/j.astropartphys.2009.01.005},
archivePrefix = {arXiv},
       eprint = {0803.2120},
 primaryClass = {astro-ph},
       adsurl = {https://ui.adsabs.harvard.edu/abs/2009APh....31..226M}
}

@ARTICLE{Fermi09,
       author = {{Abdo}, A.~A. and {Ackermann}, M. and {Arimoto}, M. and {Asano}, K. and {Atwood}, W.~B. and {Axelsson}, M. and {Baldini}, L. and {Ballet}, J. and {Band}, D.~L. and {Barbiellini}, G. and {Baring}, M.~G. and {Bastieri}, D. and {Battelino}, M. and {Baughman}, B.~M. and {Bechtol}, K. and {Bellardi}, F. and {Bellazzini}, R. and {Berenji}, B. and {Bhat}, P.~N. and {Bissaldi}, E. and {Blandford}, R.~D. and {Bloom}, E.~D. and {Bogaert}, G. and {Bogart}, J.~R. and {Bonamente}, E. and {Bonnell}, J. and {Borgland}, A.~W. and {Bouvier}, A. and {Bregeon}, J. and {Brez}, A. and {Briggs}, M.~S. and {Brigida}, M. and {Bruel}, P. and {Burnett}, T.~H. and {Burrows}, D. and {Busetto}, G. and {Caliandro}, G.~A. and {Cameron}, R.~A. and {Caraveo}, P.~A. and {Casandjian}, J.~M. and {Ceccanti}, M. and {Cecchi}, C. and {Celotti}, A. and {Charles}, E. and {Chekhtman}, A. and {Cheung}, C.~C. and {Chiang}, J. and {Ciprini}, S. and {Claus}, R. and {Cohen-Tanugi}, J. and {Cominsky}, L.~R. and {Connaughton}, V. and {Conrad}, J. and {Costamante}, L. and {Cutini}, S. and {DeKlotz}, M. and {Dermer}, C.~D. and {de Angelis}, A. and {de Palma}, F. and {Digel}, S.~W. and {Dingus}, B.~L. and {do Couto e Silva}, E. and {Drell}, P.~S. and {Dubois}, R. and {Dumora}, D. and {Edmonds}, Y. and {Evans}, P.~A. and {Fabiani}, D. and {Farnier}, C. and {Favuzzi}, C. and {Finke}, J. and {Fishman}, G. and {Focke}, W.~B. and {Frailis}, M. and {Fukazawa}, Y. and {Funk}, S. and {Fusco}, P. and {Gargano}, F. and {Gasparrini}, D. and {Gehrels}, N. and {Germani}, S. and {Giebels}, B. and {Giglietto}, N. and {Giommi}, P. and {Giordano}, F. and {Glanzman}, T. and {Godfrey}, G. and {Goldstein}, A. and {Granot}, J. and {Greiner}, J. and {Grenier}, I.~A. and {Grondin}, M. -H. and {Grove}, J.~E. and {Guillemot}, L. and {Guiriec}, S. and {Haller}, G. and {Hanabata}, Y. and {Harding}, A.~K. and {Hayashida}, M. and {Hays}, E. and {Morata}, J.~A. Hernando and {Hoover}, A. and {Hughes}, R.~E. and {J{\'o}hannesson}, G. and {Johnson}, A.~S. and {Johnson}, R.~P. and {Johnson}, T.~J. and {Johnson}, W.~N. and {Kamae}, T. and {Katagiri}, H. and {Kataoka}, J. and {Kavelaars}, A. and {Kawai}, N. and {Kelly}, H. and {Kennea}, J. and {Kerr}, M. and {Kippen}, R.~M. and {Kn{\"o}dlseder}, J. and {Kocevski}, D. and {Kocian}, M.~L. and {Komin}, N. and {Kouveliotou}, C. and {Kuehn}, F. and {Kuss}, M. and {Lande}, J. and {Landriu}, D. and {Larsson}, S. and {Latronico}, L. and {Lavalley}, C. and {Lee}, B. and {Lee}, S. -H. and {Lemoine-Goumard}, M. and {Lichti}, G.~G. and {Longo}, F. and {Loparco}, F. and {Lott}, B. and {Lovellette}, M.~N. and {Lubrano}, P. and {Madejski}, G.~M. and {Makeev}, A. and {Marangelli}, B. and {Mazziotta}, M.~N. and {McBreen}, S. and {McEnery}, J.~E. and {McGlynn}, S. and {Meegan}, C. and {M{\'e}sz{\'a}ros}, P. and {Meurer}, C. and {Michelson}, P.~F. and {Minuti}, M. and {Mirizzi}, N. and {Mitthumsiri}, W. and {Mizuno}, T. and {Moiseev}, A.~A. and {Monte}, C. and {Monzani}, M.~E. and {Moretti}, E. and {Morselli}, A. and {Moskalenko}, I.~V. and {Murgia}, S. and {Nakamori}, T. and {Nelson}, D. and {Nolan}, P.~L. and {Norris}, J.~P. and {Nuss}, E. and {Ohno}, M. and {Ohsugi}, T. and {Okumura}, A. and {Omodei}, N. and {Orlando}, E. and {Ormes}, J.~F. and {Ozaki}, M. and {Paciesas}, W.~S. and {Paneque}, D. and {Panetta}, J.~H. and {Parent}, D. and {Pelassa}, V. and {Pepe}, M. and {Perri}, M. and {Pesce-Rollins}, M. and {Petrosian}, V. and {Pinchera}, M. and {Piron}, F. and {Porter}, T.~A. and {Preece}, R. and {Rain{\`o}}, S. and {Ramirez-Ruiz}, E. and {Rando}, R. and {Rapposelli}, E. and {Razzano}, M. and {Razzaque}, S. and {Rea}, N. and {Reimer}, A. and {Reimer}, O. and {Reposeur}, T. and {Reyes}, L.~C. and {Ritz}, S. and {Rochester}, L.~S. and {Rodriguez}, A.~Y. and {Roth}, M.},
        title = "{Fermi Observations of High-Energy Gamma-Ray Emission from GRB 080916C}",
      journal = {Science},
         year = 2009,
        month = mar,
       volume = {323},
       number = {5922},
        pages = {1688},
          doi = {10.1126/science.1169101},
       adsurl = {https://ui.adsabs.harvard.edu/abs/2009Sci...323.1688A}
}

@ARTICLE{MAGICGRB,
       author = {{MAGIC Collaboration} and {Acciari}, V.~A. and {Ansoldi}, S. and {Antonelli}, L.~A. and {Arbet Engels}, A. and {Baack}, D. and {Babi{\'c}}, A. and {Banerjee}, B. and {Barres de Almeida}, U. and {Barrio}, J.~A. and {Becerra Gonz{\'a}lez}, J. and {Bednarek}, W. and {Bellizzi}, L. and {Bernardini}, E. and {Berti}, A. and {Besenrieder}, J. and {Bhattacharyya}, W. and {Bigongiari}, C. and {Biland}, A. and {Blanch}, O. and {Bonnoli}, G. and {Bo{\v{s}}njak}, {\v{Z}}. and {Busetto}, G. and {Carosi}, A. and {Carosi}, R. and {Ceribella}, G. and {Chai}, Y. and {Chilingaryan}, A. and {Cikota}, S. and {Colak}, S.~M. and {Colin}, U. and {Colombo}, E. and {Contreras}, J.~L. and {Cortina}, J. and {Covino}, S. and {D'Amico}, G. and {D'Elia}, V. and {da Vela}, P. and {Dazzi}, F. and {de Angelis}, A. and {de Lotto}, B. and {Delfino}, M. and {Delgado}, J. and {Depaoli}, D. and {di Pierro}, F. and {di Venere}, L. and {Do Souto Espi{\~n}eira}, E. and {Dominis Prester}, D. and {Donini}, A. and {Dorner}, D. and {Doro}, M. and {Elsaesser}, D. and {Fallah Ramazani}, V. and {Fattorini}, A. and {Fern{\'a}ndez-Barral}, A. and {Ferrara}, G. and {Fidalgo}, D. and {Foffano}, L. and {Fonseca}, M.~V. and {Font}, L. and {Fruck}, C. and {Fukami}, S. and {Gallozzi}, S. and {Garc{\'\i}a L{\'o}pez}, R.~J. and {Garczarczyk}, M. and {Gasparyan}, S. and {Gaug}, M. and {Giglietto}, N. and {Giordano}, F. and {Godinovi{\'c}}, N. and {Green}, D. and {Guberman}, D. and {Hadasch}, D. and {Hahn}, A. and {Herrera}, J. and {Hoang}, J. and {Hrupec}, D. and {H{\"u}tten}, M. and {Inada}, T. and {Inoue}, S. and {Ishio}, K. and {Iwamura}, Y. and {Jouvin}, L. and {Kerszberg}, D. and {Kubo}, H. and {Kushida}, J. and {Lamastra}, A. and {Lelas}, D. and {Leone}, F. and {Lindfors}, E. and {Lombardi}, S. and {Longo}, F. and {L{\'o}pez}, M. and {L{\'o}pez-Coto}, R. and {L{\'o}pez-Oramas}, A. and {Loporchio}, S. and {Machado de Oliveira Fraga}, B. and {Maggio}, C. and {Majumdar}, P. and {Makariev}, M. and {Mallamaci}, M. and {Maneva}, G. and {Manganaro}, M. and {Mannheim}, K. and {Maraschi}, L. and {Mariotti}, M. and {Mart{\'\i}nez}, M. and {Masuda}, S. and {Mazin}, D. and {Mi{\'c}anovi{\'c}}, S. and {Miceli}, D. and {Minev}, M. and {Miranda}, J.~M. and {Mirzoyan}, R. and {Molina}, E. and {Moralejo}, A. and {Morcuende}, D. and {Moreno}, V. and {Moretti}, E. and {Munar-Adrover}, P. and {Neustroev}, V. and {Nigro}, C. and {Nilsson}, K. and {Ninci}, D. and {Nishijima}, K. and {Noda}, K. and {Nogu{\'e}s}, L. and {N{\"o}the}, M. and {Nozaki}, S. and {Paiano}, S. and {Palacio}, J. and {Palatiello}, M. and {Paneque}, D. and {Paoletti}, R. and {Paredes}, J.~M. and {Pe{\~n}il}, P. and {Peresano}, M. and {Persic}, M. and {Prada Moroni}, P.~G. and {Prandini}, E. and {Puljak}, I. and {Rhode}, W. and {Rib{\'o}}, M. and {Rico}, J. and {Righi}, C. and {Rugliancich}, A. and {Saha}, L. and {Sahakyan}, N. and {Saito}, T. and {Sakurai}, S. and {Satalecka}, K. and {Schmidt}, K. and {Schweizer}, T. and {Sitarek}, J. and {{\v{S}}nidari{\'c}}, I. and {Sobczynska}, D. and {Somero}, A. and {Stamerra}, A. and {Strom}, D. and {Strzys}, M. and {Suda}, Y. and {Suri{\'c}}, T. and {Takahashi}, M. and {Tavecchio}, F. and {Temnikov}, P. and {Terzi{\'c}}, T. and {Teshima}, M. and {Torres-Alb{\`a}}, N. and {Tosti}, L. and {Tsujimoto}, S. and {Vagelli}, V. and {van Scherpenberg}, J. and {Vanzo}, G. and {Vazquez Acosta}, M. and {Vigorito}, C.~F. and {Vitale}, V. and {Vovk}, I. and {Will}, M. and {Zari{\'c}}, D. and {Nava}, L.},
        title = "{Teraelectronvolt emission from the {\ensuremath{\gamma}}-ray burst GRB 190114C}",
      journal = {\nat},
         year = 2019,
        month = nov,
       volume = {575},
       number = {7783},
        pages = {455-458},
          doi = {10.1038/s41586-019-1750-x},
archivePrefix = {arXiv},
       eprint = {2006.07249},
 primaryClass = {astro-ph.HE},
       adsurl = {https://ui.adsabs.harvard.edu/abs/2019Natur.575..455M}
}

@ARTICLE{MAGIC,
       author = {{Acciari}, V.~A. and {Ansoldi}, S. and {Antonelli}, L.~A. and {Arbet Engels}, A. and {Baack}, D. and {Babi{\'c}}, A. and {Banerjee}, B. and {Barres de Almeida}, U. and {Barrio}, J.~A. and {Becerra Gonz{\'a}lez}, J. and {Bednarek}, W. and {Bellizzi}, L. and {Bernardini}, E. and {Berti}, A. and {Besenrieder}, J. and {Bhattacharyya}, W. and {Bigongiari}, C. and {Biland}, A. and {Blanch}, O. and {Bonnoli}, G. and {Bo{\v{s}}njak}, {\v{Z}}. and {Busetto}, G. and {Carosi}, R. and {Ceribella}, G. and {Cerruti}, M. and {Chai}, Y. and {Chilingarian}, A. and {Cikota}, S. and {Colak}, S.~M. and {Colin}, U. and {Colombo}, E. and {Contreras}, J.~L. and {Cortina}, J. and {Covino}, S. and {D'Amico}, G. and {D'Elia}, V. and {da Vela}, P. and {Dazzi}, F. and {de Angelis}, A. and {de Lotto}, B. and {Delfino}, M. and {Delgado}, J. and {Depaoli}, D. and {di Pierro}, F. and {di Venere}, L. and {Do Souto Espi{\~n}eira}, E. and {Dominis Prester}, D. and {Donini}, A. and {Dorner}, D. and {Doro}, M. and {Elsaesser}, D. and {Fallah Ramazani}, V. and {Fattorini}, A. and {Ferrara}, G. and {Foffano}, L. and {Fonseca}, M.~V. and {Font}, L. and {Fruck}, C. and {Fukami}, S. and {Garc{\'\i}a L{\'o}pez}, R.~J. and {Garczarczyk}, M. and {Gasparyan}, S. and {Gaug}, M. and {Giglietto}, N. and {Giordano}, F. and {Gliwny}, P. and {Godinovi{\'c}}, N. and {Green}, D. and {Hadasch}, D. and {Hahn}, A. and {Herrera}, J. and {Hoang}, J. and {Hrupec}, D. and {H{\"u}tten}, M. and {Inada}, T. and {Inoue}, S. and {Ishio}, K. and {Iwamura}, Y. and {Jouvin}, L. and {Kajiwara}, Y. and {Karjalainen}, M. and {Kerszberg}, D. and {Kobayashi}, Y. and {Kubo}, H. and {Kushida}, J. and {Lamastra}, A. and {Lelas}, D. and {Leone}, F. and {Lindfors}, E. and {Lombardi}, S. and {Longo}, F. and {L{\'o}pez}, M. and {L{\'o}pez-Coto}, R. and {L{\'o}pez-Oramas}, A. and {Loporchio}, S. and {Machado de Oliveira Fraga}, B. and {Maggio}, C. and {Majumdar}, P. and {Makariev}, M. and {Mallamaci}, M. and {Maneva}, G. and {Manganaro}, M. and {Mannheim}, K. and {Maraschi}, L. and {Mariotti}, M. and {Mart{\'\i}nez}, M. and {Mazin}, D. and {Mender}, S. and {Mi{\'c}anovi{\'c}}, S. and {Miceli}, D. and {Miener}, T. and {Minev}, M. and {Miranda}, J.~M. and {Mirzoyan}, R. and {Molina}, E. and {Moralejo}, A. and {Morcuende}, D. and {Moreno}, V. and {Moretti}, E. and {Munar-Adrover}, P. and {Neustroev}, V. and {Nigro}, C. and {Nilsson}, K. and {Ninci}, D. and {Nishijima}, K. and {Noda}, K. and {Nogu{\'e}s}, L. and {Nozaki}, S. and {Ohtani}, Y. and {Oka}, T. and {Otero-Santos}, J. and {Palatiello}, M. and {Paneque}, D. and {Paoletti}, R. and {Paredes}, J.~M. and {Pavleti{\'c}}, L. and {Pe{\~n}il}, P. and {Perennes}, C. and {Peresano}, M. and {Persic}, M. and {Prada Moroni}, P.~G. and {Prandini}, E. and {Puljak}, I. and {Rhode}, W. and {Rib{\'o}}, M. and {Rico}, J. and {Righi}, C. and {Rugliancich}, A. and {Saha}, L. and {Sahakyan}, N. and {Saito}, T. and {Sakurai}, S. and {Satalecka}, K. and {Schleicher}, B. and {Schmidt}, K. and {Schweizer}, T. and {Sitarek}, J. and {{\v{S}}nidari{\'c}}, I. and {Sobczynska}, D. and {Spolon}, A. and {Stamerra}, A. and {Strom}, D. and {Strzys}, M. and {Suda}, Y. and {Suri{\'c}}, T. and {Takahashi}, M. and {Tavecchio}, F. and {Temnikov}, P. and {Terzi{\'c}}, T. and {Teshima}, M. and {Torres-Alb{\`a}}, N. and {Tosti}, L. and {van Scherpenberg}, J. and {Vanzo}, G. and {Vazquez Acosta}, M. and {Ventura}, S. and {Verguilov}, V. and {Vigorito}, C.~F. and {Vitale}, V. and {Vovk}, I. and {Will}, M. and {Zari{\'c}}, D. and {Nava}, L. and {MAGIC Collaboration}},
        title = "{Bounds on Lorentz Invariance Violation from MAGIC Observation of GRB 190114C}",
      journal = {\prl},
         year = 2020,
        month = jul,
       volume = {125},
       number = {2},
          eid = {021301},
        pages = {021301},
          doi = {10.1103/PhysRevLett.125.021301},
archivePrefix = {arXiv},
       eprint = {2001.09728},
 primaryClass = {astro-ph.HE},
       adsurl = {https://ui.adsabs.harvard.edu/abs/2020PhRvL.125b1301A}
}

@ARTICLE{Vaso13,
       author = {{Vasileiou}, V. and {Jacholkowska}, A. and {Piron}, F. and {Bolmont}, J. and {Couturier}, C. and {Granot}, J. and {Stecker}, F.~W. and {Cohen-Tanugi}, J. and {Longo}, F.},
        title = "{Constraints on Lorentz invariance violation from Fermi-Large Area Telescope observations of gamma-ray bursts}",
      journal = {\prd},
         year = 2013,
        month = jun,
       volume = {87},
       number = {12},
          eid = {122001},
        pages = {122001},
          doi = {10.1103/PhysRevD.87.122001},
archivePrefix = {arXiv},
       eprint = {1305.3463},
 primaryClass = {astro-ph.HE},
       adsurl = {https://ui.adsabs.harvard.edu/abs/2013PhRvD..87l2001V}
}

@ARTICLE{Du20,
       author = {{Du}, Shen-Shi and {Lan}, Lin and {Wei}, Jun-Jie and {Zhou}, Zi-Min and {Gao}, He and {Jiang}, Lu-Yao and {Zhang}, Bin-Bin and {Liu}, Zi-Ke and {Wu}, Xue-Feng and {Liang}, En-Wei and {Zhu}, Zong-Hong},
        title = "{Lorentz Invariance Violation Limits from the Spectral-lag Transition of GRB 190114C}",
      journal = {\apj},
         year = 2021,
        month = jan,
       volume = {906},
       number = {1},
          eid = {8},
        pages = {8},
          doi = {10.3847/1538-4357/abc624},
archivePrefix = {arXiv},
       eprint = {2010.16029},
 primaryClass = {astro-ph.HE},
       adsurl = {https://ui.adsabs.harvard.edu/abs/2021ApJ...906....8D}
}
\section*{Appendix : Tests with synthetic data}
\begin{figure}[H]
    \centering
    \includegraphics[width=0.7\linewidth]{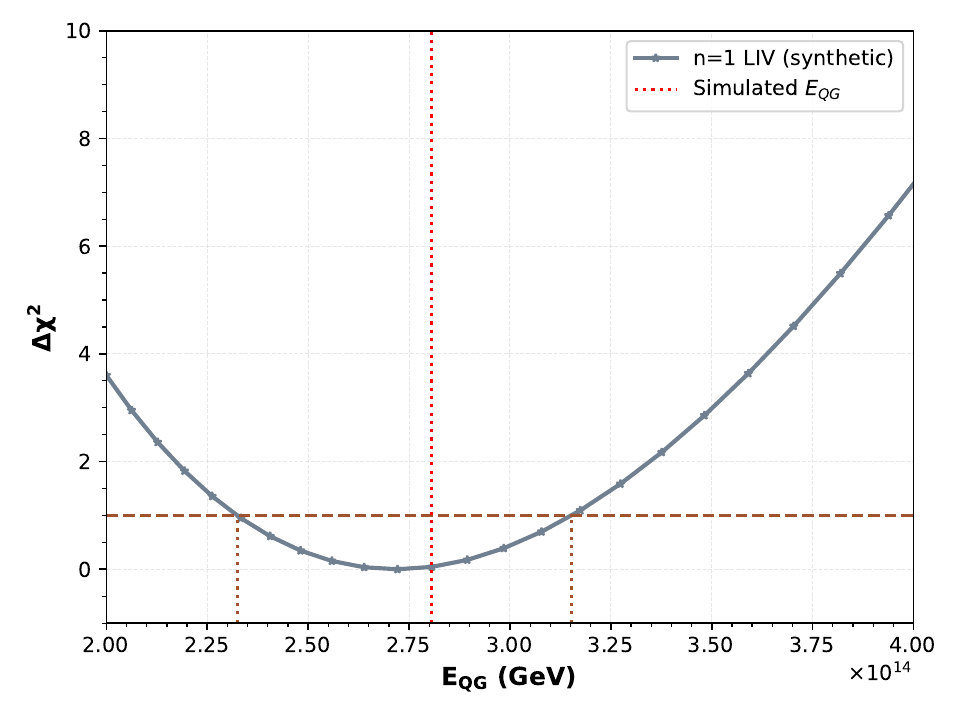}
    \caption{\rthis{$\Delta \chi^2$, defined as ($\chi^2-\chi^2_{min}$), plotted against $E_{QG}$ for a linearly-dependent LIV, obtained from generating synthetic data samples. The input value of $E_{QG}$ used to generate the synthetic samples, is the best-fit obtained earlier in Fig.\ref{fig:linearLIV}, $E_{QG}=2.81 \times 10^{14}$. The horizontal dashed line represents $\Delta\chi^2 = 1$ and gives us the 68.3\% confidence interval for $E_{QG,1,synt} = 2.72^{+0.43}_{-0.40}\times 10^{14}$ GeV.}}
    \label{fig:linearLIV_synt}
\end{figure}

\begin{figure}[H]
    \centering
    \includegraphics[width=0.7\linewidth]{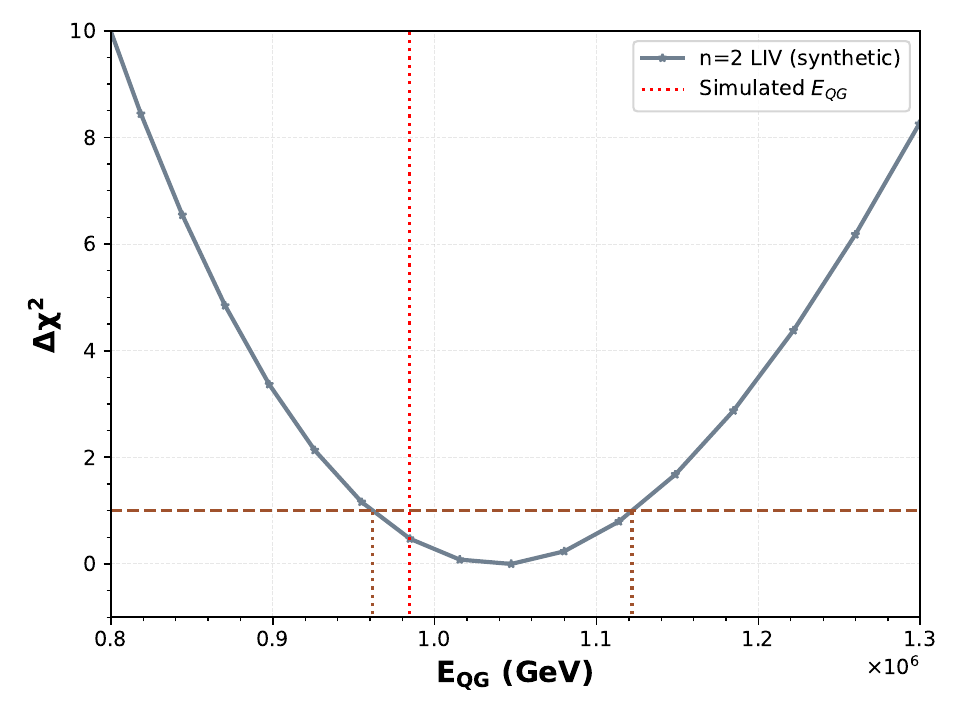}
    \caption{\rthis{$\Delta \chi^2$, defined as ($\chi^2-\chi^2_{min}$), plotted against $E_{QG}$ for a quadratically-dependent LIV, obtained from generating synthetic data samples. The input value of $E_{QG}$ used to generate the synthetic samples, is the best-fit obtained earlier in Fig.\ref{fig:quadLIV}, $E_{QG}=9.85 \times 10^{5}$. The horizontal dashed line represents $\Delta\chi^2 = 1$ and gives us the 68.3\% confidence interval for $E_{QG,2,synt} = 1.05^{+0.07}_{-0.09}\times 10^{6}$ GeV.}}
    \label{fig:quadLIV_synt}
\end{figure}

\rthis{We now generate  a single  realization of synthetic data and test whether we can recover the best-fit parameters using the profile likelihood technique described in the main manuscript. For this purpose, we generated synthetic spectral lag  data consisting of a  superposition of both LIV and intrinsic astrophysical emission (using Eq.~\ref{eq:sum}). We then added uncertainties to each lag datum, which are  generated from a normal distribution with mean 0 and standard deviation equal to the input uncertainty for that point. The input $E_{QG}$ values  are given by  $E_{QG}=2.81 \times 10^{14}$ GeV  (linear LIV) and  $E_{QG}=9.85 \times 10^{5}$ GeV, respectively, which is the minimum value we got for the real data. We then followed the same procedure  to obtain the best-fit values for $E_{QG}$ as for the real data. Recovery of $E_{QG}$ using these synthetic data can be found in Fig.~\ref{fig:linearLIV_synt} and Fig.~\ref{fig:quadLIV_synt}, for linear and quadratic LIV, respectively.  The best-fit recovered values are: $E_{QG} =  2.72^{+0.43}_{-0.40}\times 10^{14}$ GeV (linear LIV) and $E_{QG} =  1.05^{+0.07}_{-0.09 }\times 10^{6}$ GeV (quadratic LIV).  Therefore, in both cases, we can recover the input value of $E_{QG}$ for these synthetic data using the profile likelihood technique. }

\end{document}